\RequirePackage{fix-cm}
\documentclass{svjour3}                     % onecolumn (standard format)
\smartqed  % flush right qed marks, e.g. at end of proof
\usepackage{graphicx}
\begin{document}

\title{
Recent results and future prospects of kaonic nuclei at J-PARC
}


\author{
F.Sakuma \and
S.Ajimura \and
T.Akaishi \and
H.Asano \and
M.Bazzi \and
G.Beer \and
H.Bhang \and
M.Bragadireanu \and
P.Buehler \and
L.Busso \and
M.Cargnelli \and
S.Choi \and
A.Clozza \and
C.Curceanu \and
S.Enomoto \and
H.Fujioka \and
Y.Fujiwara \and
T.Fukuda \and
C.Guaraldo \and
T.Hashimoto \and
R.S.Hayano \and
T.Hiraiwa \and
M.Iio \and
M.Iliescu \and
K.Inoue \and
Y.Ishiguro \and
T.Ishikawa \and
S.Ishimoto \and
K.Itahashi \and
M.Iwasaki \and
M.Iwai \and
K.Kanno \and
K.Kato \and
Y.Kato \and
S.Kawasaki \and
P.Kienle \and
H.Kou \and
Y.Ma \and
J.Marton \and
Y.Matsuda \and
M.Miliucci \and
Y.Mizoi \and
O.Morra \and
R.Murayama \and
T.Nagae \and
H.Noumi \and
H.Ohnishi \and
S.Okada \and
H.Outa \and
K.Ozawa \and
K.Piscicchia \and
Y.Sada \and
A.Sakaguchi \and
M.Sato \and
A.Scordo \and
M.Sekimoto \and
H.Shi \and
K.Shirotori \and
M.Simon \and
D.Sirghi \and
F.Sirghi \and
S.Suzuki \and
T.Suzuki \and
K.Tanida \and
H.Tatsuno \and
M.Tokuda \and
D.Tomono \and
A.Toyoda \and
K.Tsukada \and
O.V\'azquez Doce \and
E.Widmann \and
T.Yamaga \and
T.Yamazaki \and
C.Yoshida \and
Q.Zhang \and
J.Zmeskal
}


\institute{F.Sakuma, H.Asano, K.Itahashi, M.Iwasaki, K.Kanno, Y.Kato, Y.Ma, R.Murayama, H.Outa, T.Yamaga, T.Yamazaki, Q.Zhang \at
RIKEN Cluster for Pioneering Research, RIKEN, Wako 351-0198, Japan
%              Tel.: +123-45-678910\\
%              Fax: +123-45-678910\\
              \email{sakuma@ribf.riken.jp}           %  \\
           \and
           S.Ajimura, T.Hiraiwa, K.Inoue, S.Kawasaki, H.Noumi, K.Shirotori, D.Tomono \at
Research Center for Nuclear Physics (RCNP), Osaka University, Osaka 567-0047, Japan
           \and
           T.Akaishi, A.Sakaguchi \at
Department of Physics, Osaka University, Osaka 560-0043, Japan
           \and
           G.Beer \at
Department of Physics and Astronomy, University of Victoria, Victoria, British Columbia V8W 3P6, Canada
           \and
           H.Bhang, S.Choi \at
Department of Physics, Seoul National University, Seoul 151-742, South Korea
           \and
           M.Bragadireanu, D.Sirghi, F.Sirghi \at
National Institute of Physics and Nuclear Engineering - IFIN HH, Bucharest, Romania
           \and
           P.Buehler, M.Cargnelli, J.Marton, H.Shi, M.Simon, E.Widmann, J.Zmeskal \at
Stefan-Meyer-Institut f\"{u}r subatomare Physik, A-1020 Vienna, Austria
           \and
           L.Busso, O.Morra \at
Istituto Nazionale di Fisica Nucleare (INFN), Sezione di Torino, Torino, Italy
           \and
           L.Busso \at
Dipartimento di Fisica Generale, Universita' di Torino, Torino, Italy
           \and
           M.Bazzi, A.Clozza, C.Curceanu, C.Guaraldo, M.Iliescu, M.Miliucci, K.Piscicchia, A.Scordo, D.Sirghi, 
F.Sirghi, O.V\'azquez Doce \at
Laboratori Nazionali di Frascati dell' INFN, I-00044 Frascati, Italy
           \and
           S.Enomoto, M.Iio, S.Ishimoto, M.Iwai, H.Noumi, K.Ozawa, M.Sato, M.Sekimoto, S.Suzuki, A.Toyoda \at
High Energy Accelerator Research Organization (KEK), Tsukuba 305-0801, Japan
           \and
           H.Fujioka, H.Kou, M.Tokuda \at
Department of Physics, Tokyo Institute of Technology, Tokyo 152-8551, Japan
           \and
           Y.Fujiwara, R.S.Hayano, T.Ishikawa, T.Suzuki, T.Yamazaki \at
Department of Physics, The University of Tokyo, Tokyo 113-0033, Japan
           \and
           T.Fukuda, Y.Mizoi \at
Laboratory of Physics, Osaka Electro-Communication University, Osaka 572-8530, Japan
           \and
           T.Hashimoto, K.Tanida \at
ASRC, Japan Atomic Energy Agency, Ibaraki 319-1195, Japan
           \and
           Y.Ishiguro, K.Kato, T.Nagae \at
Department of Physics, Kyoto University, Kyoto 606-8502, Japan
           \and
           P. Kienle$^{\dagger}$, O.V\'azquez Doce \at
Technische Universit\"{a}t M\"{u}nchen, D-85748, Garching, Germany
           \and
           Y.Matsuda \at
Graduate School of Arts and Sciences, The University of Tokyo, Tokyo 153-8902, Japan
           \and
           H.Ohnishi, Y.Sada, C.Yoshida  \at
Research Center for Electron Photon Science (ELPH), Tohoku University, Sendai 982-0826, Japan
           \and
           S.Okada \at
Engineering Science Laboratory, Chubu University, Aichi 487-8501, Japan
           \and
           K.Piscicchia \at
Centro Ricerche Fermi-Museo Storico della Fisica e Centro studi e ricerche ``Enrico Fermi'', 000184 Rome, Italy
           \and
           H.Tatsuno \at
Department of Physics, Tokyo Metropolitan University, Tokyo 192-0397, Japan
           \and
           K.Tsukada \at
Institute for Chemical Research, Kyoto University, Kyoto 611-0011, Japan
}
\date{Received: date / Accepted: date}

\maketitle

\begin{abstract}
$\bar K$-nuclear bound systems, kaonic nuclei, have been widely
discussed as products of the strongly attractive $\bar K N$
interaction in $I = 0$ channels.
Recently, we demonstrated that kaonic nuclei can be produced via 
in-flight $(K^-,N)$ reactions using the low-momentum DC kaon beam at the
J-PARC E15 experiment.
We observed the simplest kaonic nuclei, $K^-pp$, having a much deeper
binding energy than normal nuclei.
For further studies, we have proposed a series of experimental programs
for the systematic investigation of light kaonic nuclei, from 
$\bar K N$ ($\Lambda(1405)$) to $\bar K NNNN$.
In the new experiment approved as J-PARC E80, we will measure
the $\bar K NNN$ ($A=3$) system as a first step toward a comprehensive
study. 
\end{abstract}

\section{Introduction}
\label{intro}

The study of the $\bar K N$ interaction is one of the most important
subjects in meson--baryon interactions in low energy quantum
chromodynamics (QCD).
Extensive measurements of anti-kaonic hydrogen
atoms~\cite{Iwasaki:1997wf,Beer:2005qi,Bazzi:2011zj}
and low-energy $\bar K N$ scattering~\cite{Martin:1980qe} have revealed
the strongly attractive nature of the $\bar K N$ interaction in the
isospin I = 0 channel.
Consequently, the possible existence of deeply bound kaonic nuclear
states (kaonic nuclei) has been widely
discussed~\cite{Nogami:1963xqa,Akaishi:2002bg,Yamazaki:2002uh,Shevchenko:2006xy,Shevchenko:2007ke,Ikeda:2007nz,Dote:2008in,Ikeda:2008ub,Wycech:2008wf,Dote:2008hw,Ikeda:2010tk,Barnea:2012qa,Bayar:2012hn,Maeda:2013,Revai:2014twa,Dote:2014via,Sekihara:2016vyd,Ohnishi:2017uni,Dote:2017veg,Dote:2017wkk}.
Kaonic nuclei are predicted to be compact due to the
strong $\bar K N$ attraction, suggesting that
high-density nuclear matter is realized in kaonic systems.

Among kaonic nuclei, the $\bar K NN$ system with $I = 1/2$
and $J^{P} = 0^-$ (symbolically denoted as $K^-pp$ for the $I_z = +1/2$ state) is of special
interest because it is the lightest $S = -1$ $\bar K$ nucleus.
Despite considerable experimental efforts over the past 20 years, it has been challenging to prove the existence of $K^-pp$.
Several groups have reported observations of a $K^-pp$ candidate with a
binding energy of around 100 MeV in experiments measuring
non-mesonic decay branches of $\Lambda p$ and/or $\Sigma^0 p$ in different
reactions~\cite{Agnello:2005qj,Yamazaki:2010mu,Ichikawa:2014ydh}.
There are also contradicting reports concluding that the
reactions can be understood without a bound
state~\cite{Doce:2015ust,DelGrande:2018sbv,Agakishiev:2014dha,Tokiyasu:2013mwa}.

Recently, we confirmed the existence of the $K^-pp$ bound state by using
the simplest reaction of in-flight $^3$He$(K^-,N)$ at the J-PARC E15
experiment~\cite{Hashimoto:2014cri,Sada:2016nkb,Ajimura:2018iyx,J-PARCE15:2020gbh}.
A distinct peak structure was observed well below the mass threshold of
$K^- + p + p$ in the $\Lambda p$ invariant-mass (IM) spectrum, obtained from
the $^3$He$(K^-,\Lambda p)n$ measurement.
The simplest and most natural interpretation of this peak is $K^-pp$.
This result is experimentally solid compared to previously reported
results.

To obtain further detailed information on kaonic nuclei, we have
planned a series of experimental programs using the $(K^-,N/d)$ reaction
on light nuclear targets.
The programs will enable a detailed study of a range of nuclei from $\bar K N$
($\Lambda(1405)$) to $\bar K NNNN$ using the world's highest
intensity low-momentum kaon beam at J-PARC.
The programs comprise:
\begin{itemize}
 \item Precise measurements of the $\Lambda(1405)$ state in a large
       momentum transfer region via the $d(K^-,n)$ reaction,
       to experimentally clarify whether it is a baryonic state
       or a $\bar K N$ molecular state,
 \item Investigations of the spin and parity of the $\bar K NN$ state
       via $^3$He$(K^-,N)$ reactions,
 \item A search for $\bar K NNN$ states via $^4$He$(K^-,N)$
       reactions, as a bridge to access heavier systems, and
 \item An advanced search for $\bar K NNNN$ states via the
       $^6$Li$(K^-,d)$ reaction.
\end{itemize}
In parallel to these studies, we also intend to access the $S=-2$ kaonic
nuclei, such as the theoretically predicted $K^-K^-pp$ state.
The $S=-2$ system could allow us to access even
higher density systems than the $S=-1$ kaonic nuclei.
As described in our Letter of Intent~\cite{KKpp_LoI},
one possible approach for the measurements at J-PARC could be:
\begin{itemize}
 \item Searching for $\bar K \bar K NN$ states via $\bar p {^3}$He
 annihilation.
\end{itemize}

To ensure the measurements are systematic and precise, we are planning to construct
a totally new $4\pi$ spectrometer to measure all the particles involved in
the reactions and to reconstruct their formation and decay exclusively.
The spectrometer is designed to be highly versatile so that all the
experiments can be performed simply by changing the target materials.
In addition, for more efficient use of the high-intensity kaon beam,
we have proposed shortening the existing K1.8BR beam line for a larger kaon
yield without deteriorating the momentum resolution of the kaon beam.

In this article, we review previous $K^-pp$ measurements
at the J-PARC E15 experiment in Sec.~\ref{sec:1} and set out what we have
learned from the experimental investigations dedicated to kaonic
nuclei.
In Sec.\ref{sec:2}, we present the details of the new experiment,
approved as J-PARC E80, that focuses on searching for $\bar K NNN$.

\section{Results of the J-PARC E15 Experiment}
\label{sec:1}

We conducted an experimental investigation of the $K^-pp$ bound
state using the simplest $\bar K$ induced reaction of $K^- + {^3}$He, 
via the nucleon knock-out reaction $K^- N \to \bar K n$ followed by
the two-nucleon absorption $\bar K + NN \to$ $K^-pp$.
In the experiment, we used a kaon momentum of 1 GeV/$c$, around which the
$K^- N \to \bar K n$ reactions have the maximum cross section.
The recoiled kaon $\bar K$ at a momentum $q$ behaves as an
`off-shell particle' (the total energy can be lower than its intrinsic mass)
within a time range allowed by the uncertainty principle.
The momentum transfer $q$ is defined as that between the incident kaon and the
outgoing neutron in the laboratory frame $q = | p^{lab}_{K^-} - p^{lab}_n|$.
In the reaction, we used the low-momentum back-scattered kaon as an
off-shell kaon source and the residual spectator nucleons $NN$ as an
`actual target' to form a $K^-pp$ state with an energy below the
intrinsic mass of $M(Kpp)$ (= $m_K+2m_N$ = 2.37 GeV/$c^2$).

\begin{figure}[tb]\begin{center}
 \begin{minipage}{0.45\hsize}
  \begin{center}
   \includegraphics[width=0.85\textwidth,keepaspectratio]{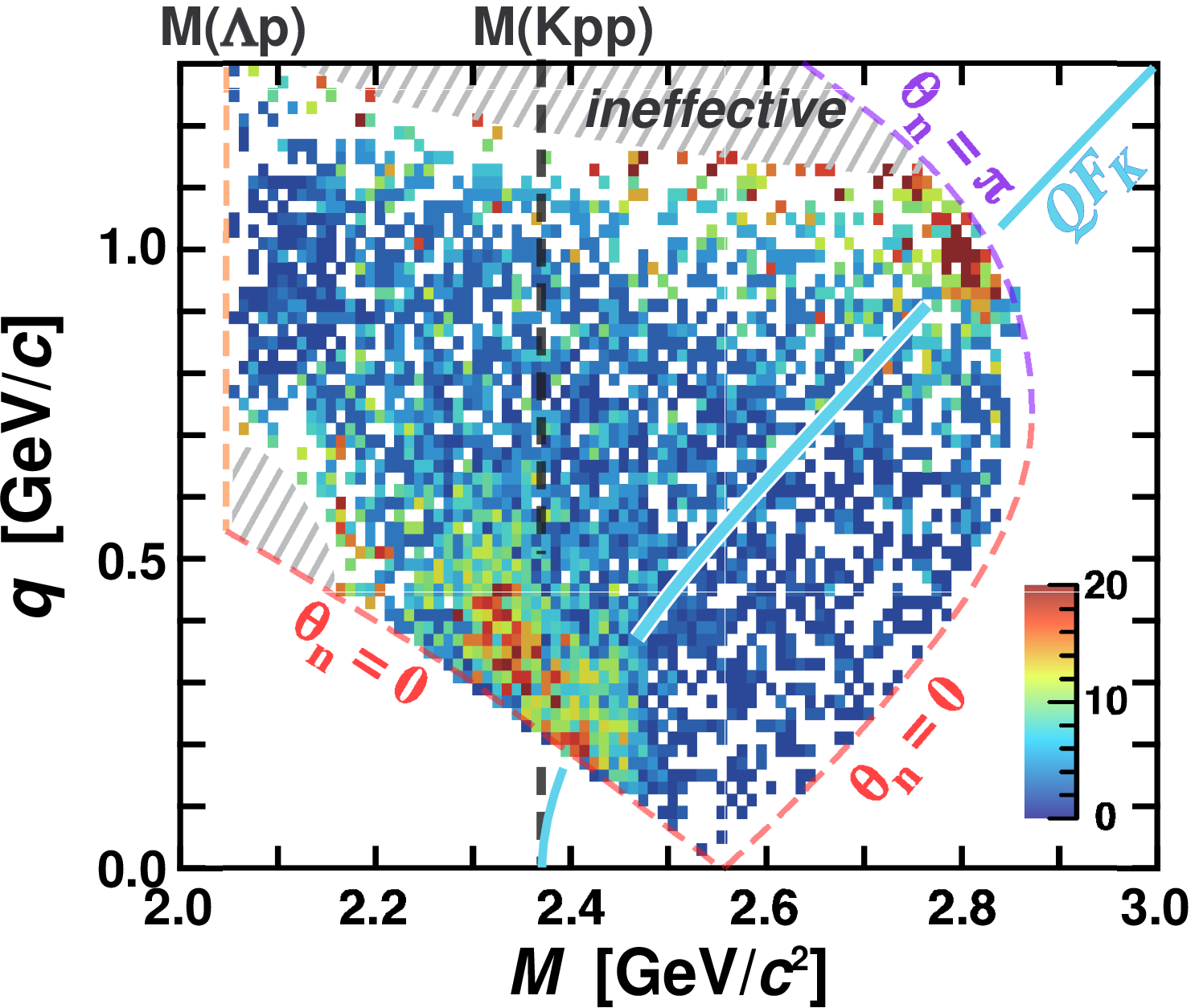}
  \end{center}
 \end{minipage}
 \begin{minipage}{0.45\hsize}
  \begin{center}
   \includegraphics[width=0.85\textwidth,keepaspectratio]{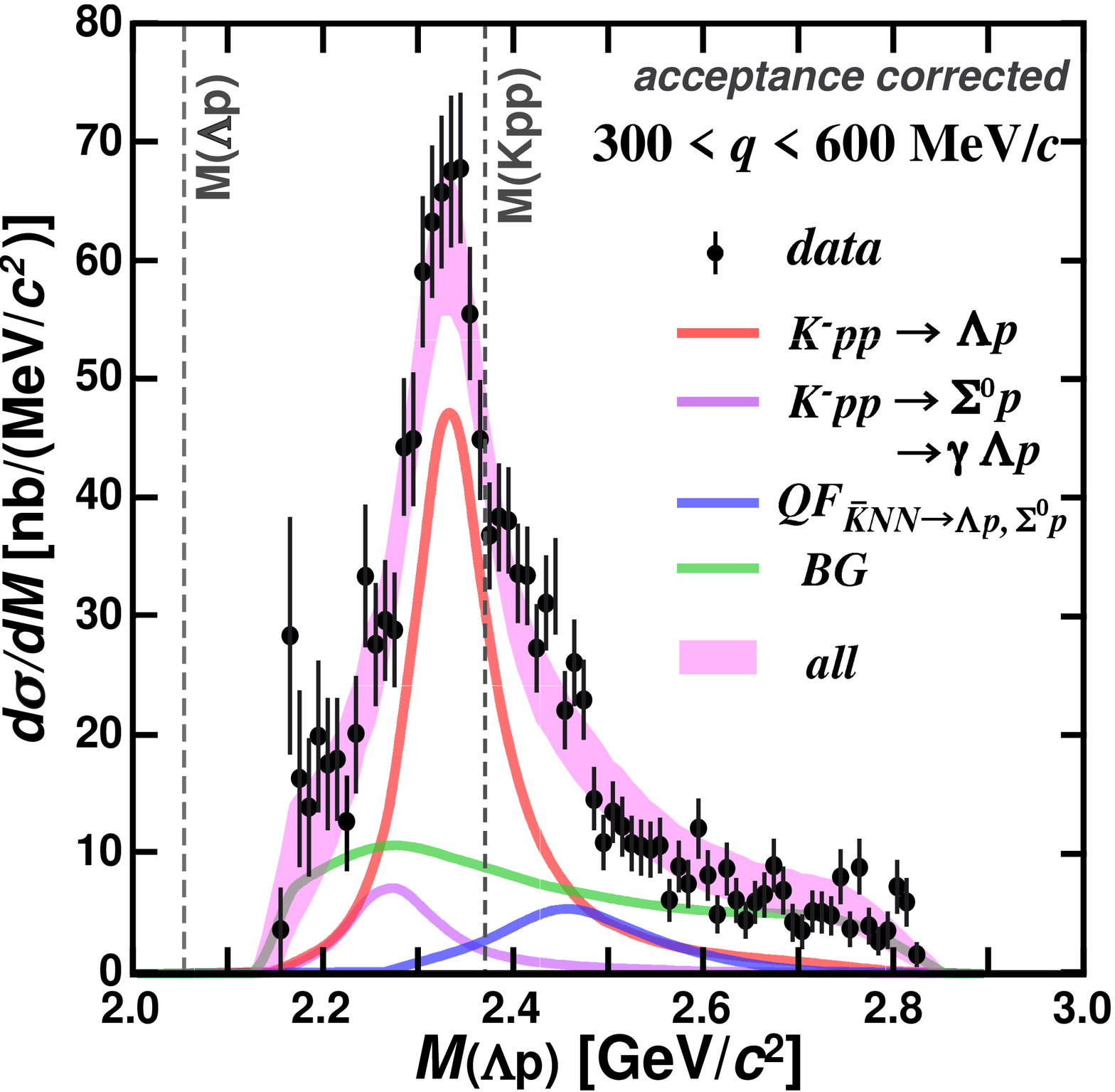}
  \end{center}
 \end{minipage}
 \caption{
 (left) Efficiency and acceptance corrected data over $IM$ ($\Lambda
 p$ invariant mass) and $q$ (momentum transfer).
 (right) $\Lambda p$ invariant mass in the region $0.3 < q < 0.6$
 MeV/$c$.
 }
\label{fig:PLB}
\end{center}
 \end{figure}

For the $\Lambda pn$ final states, we observed a kinematic anomaly in
the $\Lambda p$ invariant mass near the mass threshold of $M(Kpp)$ at
$q \sim$ 0.4
GeV/$c$~\cite{Sada:2016nkb,Ajimura:2018iyx,J-PARCE15:2020gbh}.
As shown in Fig.~\ref{fig:PLB} (left), we confirmed the existence of the
bound state below $M(Kpp)$, whose mass centroid
is independent of $q$, at a binding energy as deep as $\sim$ 50
MeV.
The back-scattered `on-shell' kaon, whose total kaon energy is above its
intrinsic mass, can also be absorbed by the spectator nucleons without
forming a bound state.
The kinematical centroid of this quasi-free absorption process is
plotted in the figure, denoted as $QF_K$.
Along the line $QF_K$, there are two event concentration points at $\theta_n =
0$ and $\theta_n = \pi$, but both are well separated from the region of
interest.
Figure~\ref{fig:PLB} (right) shows the $\Lambda p$ invariant mass spectrum
after the acceptance and efficiency corrections for the region $0.3 < q < 0.6$
GeV/$c$ where the $K^-pp$ bound state is dominant.
A clear peak originating from $K^-pp$ can be seen below
$M(Kpp)$, shown by the red line, whose binding energy reaches $\sim$ 50 MeV.

The simplest and most natural interpretation is the kaon-nuclear
bound state $K^-pp$.
The observed large form factor of $\sim$ 400 MeV/$c$, based on
a simple plane wave impulse approximation (PWIA)~\cite{Sada:2016nkb,Ajimura:2018iyx,J-PARCE15:2020gbh},
and the large binding energy of
the $K^-pp$ state imply the formation of a fairly compact and dense
system.
The observed structure below the mass threshold in the $\Lambda p$
spectrum has also been interpreted as the $\bar K NN$ quasi-bound system:
a theoretical calculation in Ref.~\cite{Sekihara:2016vyd,Sekihara:QNP2018}
demonstrated that the experimental spectrum can be reproduced with the $\bar
K NN$ quasi-bound system and the quasi-free processes based on 
a theoretical treatment of the $^3$He$(K^-,\Lambda p)n$ reaction.

Thus, the E15 experiment opened a new era of experimental research on
kaonic nuclei with the virtual kaon beam produced from in-flight
$(K^-,N)$ reactions.
This has been realized by using the world's highest intensity kaon beam
available at J-PARC.
However, we encountered several issues with the existing setup that made
further investigations of the kaonic nuclei difficult.
The new series of experiments will continue the systematic
research into light kaonic nuclei using a new large acceptance
detector system and the improved K1.8BR beam line.

\section{New Experiment (J-PARC E80 Experiment)}
\label{sec:2}

\begin{figure}[tbp]
 \begin{center}
  \includegraphics[width=0.4\textwidth,keepaspectratio]{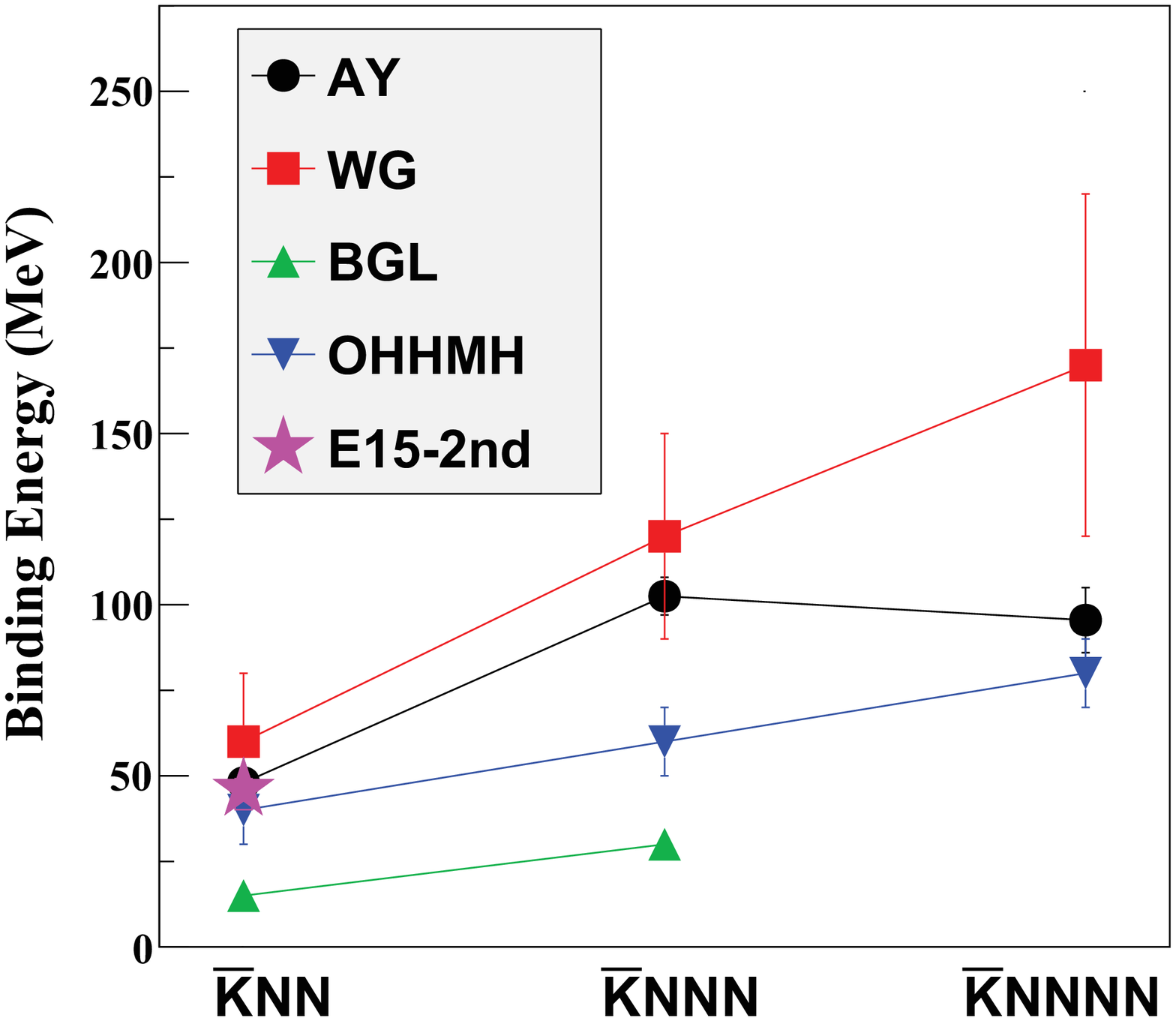}
  \includegraphics[width=0.4\textwidth,keepaspectratio]{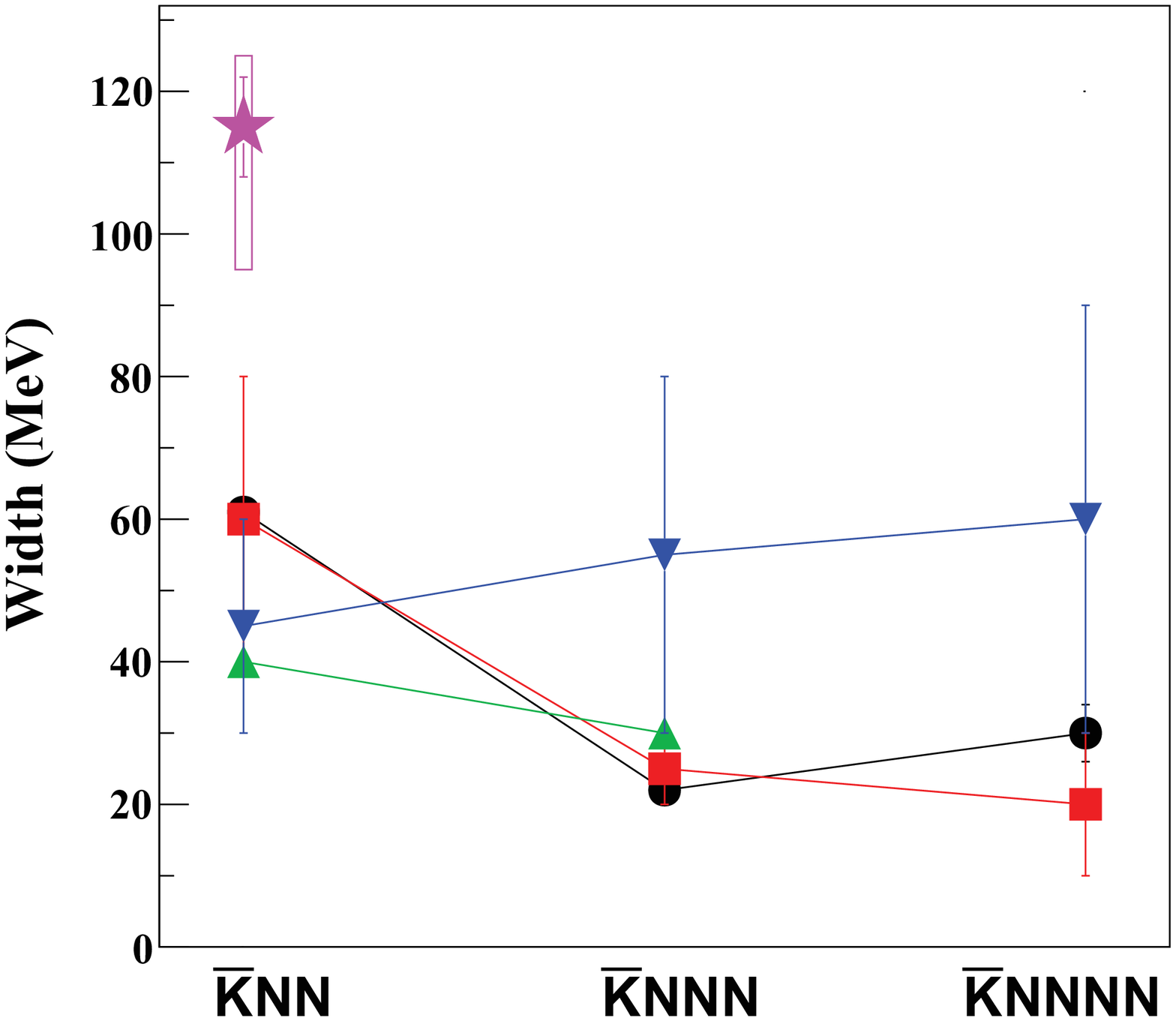}
  \caption{Summary of theoretical calculations of the kaonic nuclei
  from $A=2$ to $4$ in different models
  AY~\cite{Akaishi:2002bg,Yamazaki:2002uh},
  WG~\cite{Wycech:2008wf}, BGL~\cite{Barnea:2012qa}, and
  OHHMH~\cite{Ohnishi:2017uni}.
  The result obtained at the E15 experiment is also
  plotted~\cite{Ajimura:2018iyx}.
  }
  \label{fig:kaonic}
 \end{center}
\end{figure}

For the next stage, we aim to determine the mass number dependence of the
binding energy, decay width, and system size beyond $\bar K NN$.
The mass number dependence has been calculated with several theoretical
models, as summarized in Fig.~\ref{fig:kaonic}.
The values of the binding energy and decay width predicted by the models vary
widely due to the differences in the $\bar KN$ interaction models.
However, almost all the predictions show that the larger nuclei have
stronger binding energies. 
For the width, the theoretical calculations take into account only
mesonic decay channels, such as $\pi\Sigma N$ and $\pi\Lambda N$.
The calculated width is expected to be larger if the models adopt
non-mesonic decay channels, as demonstrated in Ref.~\cite{Bayar:2012hn}.

From an experimental point of view, when the mass number is large,
it becomes difficult to handle the number of particles in the final state
and to deduce the physics behind the reaction.
Thus, we take a step-by-step approach.
In the new experiment, J-PARC E80, we aim to measure
the $\bar K NNN$ ($A=3$)
system as a first step toward a comprehensive study. 
From the experience of E15, we have learned that
reducing the number of particles in the final state is a key to
removing ambiguity in interpreting the reaction process.
Therefore, we focus on the $K^-ppn \to \Lambda d$ and $\Lambda pn$ decay
channels in $^4$He$(K^-,\Lambda d / \Lambda pn)n$ reactions.

The new experiment will provide the mass number dependence of the kaonic nuclei
for the first time.
The dependence can more clearly reveal the $\bar KN$ interaction below the mass
threshold, by comparing the obtained properties of the $\bar K NNN$
state with those of the already reported $K^-p$ ($\Lambda(1405)$) and
$K^-pp$ states.

%---------------------------------------------------%
\subsection{\label{sec:setup}Experimental Method}
%---------------------------------------------------%
To date, stopped $K^-$ reactions have mainly been used to search for the $\bar K NNN$ state.
The KEK-PS E471/E549 collaborations measured the inclusive
$^4$He$(K^-_{stopped}, p/n)$ reactions with a spectrometer dedicated to
TOF measurement of protons and neutrons~\cite{Sato:2007sb,Yim:2010zza}.
They found no specific peak structures below the mass threshold
of $M(\bar K NNN)$ in the missing mass spectra from the reactions.
This is due to the huge background originating from the two-nucleon
absorption processes, $\bar KNN \to YN$, and quasi-free hyperon
productions and its decays, $\bar K N \to \pi Y$, whose production
mechanisms are quite complicated.
The background cannot be discriminated kinematically from the
signal with small production cross section in the inclusive measurements.

Indeed, our inclusive analysis of the in-flight $^3$He$(K^-,n)X$
measurement showed no significant peak structures due to the huge
background from quasi-free processes and the two-nucleon absorption
processes~\cite{Hashimoto:2014cri}.
By reducing the background by the exclusive measurement of
$^3$He$(K^-,\Lambda p)n$ and identifying all the final-state particles in
a wide momentum-transfer region, the $K^-pp$ signal could be
separated out.
We found that the $K^-pp$ signal has a much smaller cross section ($\sim$ 10 $\mu$b)
than the quasi-free processes ($\sim$ 10 mb) and is distributed
up to $q \sim$ 600 MeV/$c$~\cite{Sada:2016nkb,Ajimura:2018iyx,J-PARCE15:2020gbh}.

On the other hand, there are two reports of observations of the
$K^-ppn$ candidate below the mass threshold in the $\Lambda d$
invariant mass spectrum with stopped $K^-$ reactions (FINUDA) and
heavy-ion collisions (FOPI).
The FINUDA collaboration reported that the candidate has a binding
energy of $\sim$ 60 MeV and a width of $\sim$ 40 MeV~\cite{Agnello:2007aa}.
The candidate reported by the FOPI collaboration has a much deeper
binding energy of $\sim$ 150 MeV with a broader width of $\sim$ 100
MeV~\cite{Herrmann:2005}.
However, because these measurements were performed only inclusively,
possible contributions from multi-nucleon absorption processes and
$N^*/Y^*$ decays to the peak structure cannot be excluded.
Furthermore, the statistics reported were very limited and thus the
results remain speculative.

Therefore, the key to the experimental search is to adopt a simple
reaction and to measure it exclusively.
Adopting a simple reaction, such as in-flight $\bar K$ induced reactions
with light target nuclei, enables us to specify the reaction channel
using the momentum-transfer dependence.
Exclusive measurements are crucial for distinguishing small and broad
signals from the large and widely distributed quasi-free and multi-nucleon
absorption background.

In the new experiment, we will perform exclusive measurements of the
production and decay of the $K^-ppn$ state using the in-flight reaction
\begin{eqnarray*}
 K^- + ^4{\rm He} &\to& K^-ppn + n
\end{eqnarray*}
followed by the expected no-mesonic decays
\begin{eqnarray*}
 K^-ppn &\to& \Lambda + d {\rm , }\\
 K^-ppn &\to& \Lambda + p + n.
\end{eqnarray*}
We aim to determine the binding energy and width from the invariant
mass reconstruction of the decays.
The invariant mass will be obtained as a function of the momentum transfer to
distinguish the bound-state production from the quasi-free processes and
multi-nucleon absorption processes by the event kinematics as
demonstrated in the E15 analysis.
In the $K^- + {^4}{\rm He}$ reaction, it is possible
to measure the isospin partner of $K^-ppn$, {\it i.e.}, the
$K^-pnn$ state via
\begin{eqnarray*}
 K^- + ^4{\rm He} &\to& K^-pnn + p, \\
  K^-pnn &\to& \Lambda + n + n.
\end{eqnarray*}
Comparing the properties of the isospin partners is of special
importance for investigating the internal composition of the kaonic nuclei.
On the other hand, the measurement is challenging,
because two-neutron detection is required to identify the $K^-pnn$
decay.

%---------------------------------------------------%
\subsection{\label{sec:setup}Apparatus}
%---------------------------------------------------%

\begin{figure}[tbp]
 \begin{center}
  \includegraphics[width=0.7\textwidth,keepaspectratio]{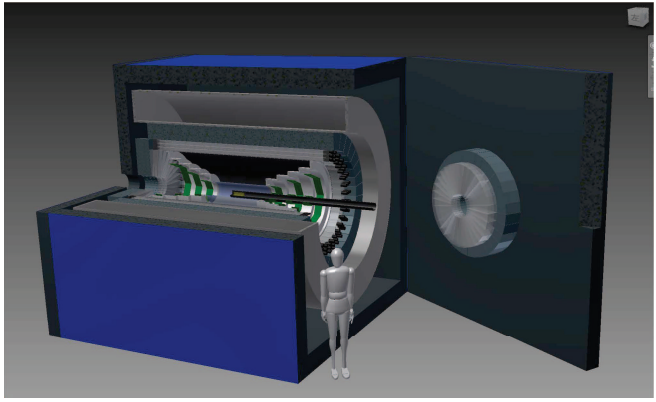}
  \caption{
  Conceptual design of the CDS.
  }
  \label{fig:CDS_3D}
 \end{center}
\end{figure}

We aim to produce the $\bar K NNN$ state using the $(K^-,N)$
reaction at the K1.8BR beam line, as successfully performed in the
previous experiment E15; a recoiled virtual kaon ($\bar K$) generated
by $K^-N \to \bar K N$ processes can be directly induced into 
residual nucleons within the strong interaction range.
We will utilize 1.0 GeV/$c$ incident kaons to maximize the $\bar K N$ reaction
rate at around zero degrees.
The incoming $K^-$ beam will be identified and its momentum analyzed by the
beam-line spectrometer.
The beam kaon will irradiate a liquid $^4$He target located at the final
focus point, and all the particles generated from the reactions will be
identified with a cylindrical detector system (CDS) surrounding the target system.
A conceptual design of the CDS is shown in Fig.~\ref{fig:CDS_3D}. It is
mainly composed of a large superconducting solenoid magnet, a cylindrical
wire drift chamber, and a cylindrical neutron detector.
The kaonic nuclei will then be identified via invariant-mass reconstruction 
of the decay particles.
By detecting the nucleon coming from the initial $(K^-,N)$ reaction, or
by identifying it with the missing mass technique, we will realize
exclusive measurement of the kaonic nuclei.
The details of the apparatus used for the experiment can be found in
Ref.~\cite{E80_proposal}.

%---------------------------------------------------%
\subsection{Expected Spectrum}
%---------------------------------------------------%

\begin{figure}[tbp]
 \begin{center}
  \includegraphics[width=0.8\textwidth,keepaspectratio]{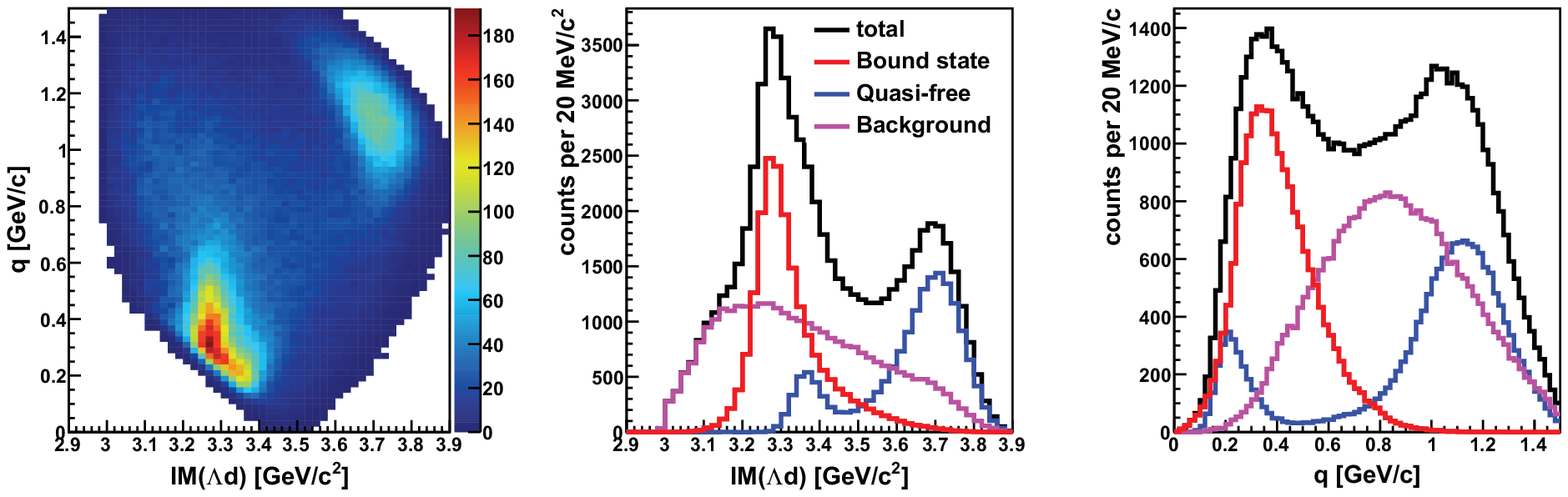}
  \includegraphics[width=0.8\textwidth,keepaspectratio]{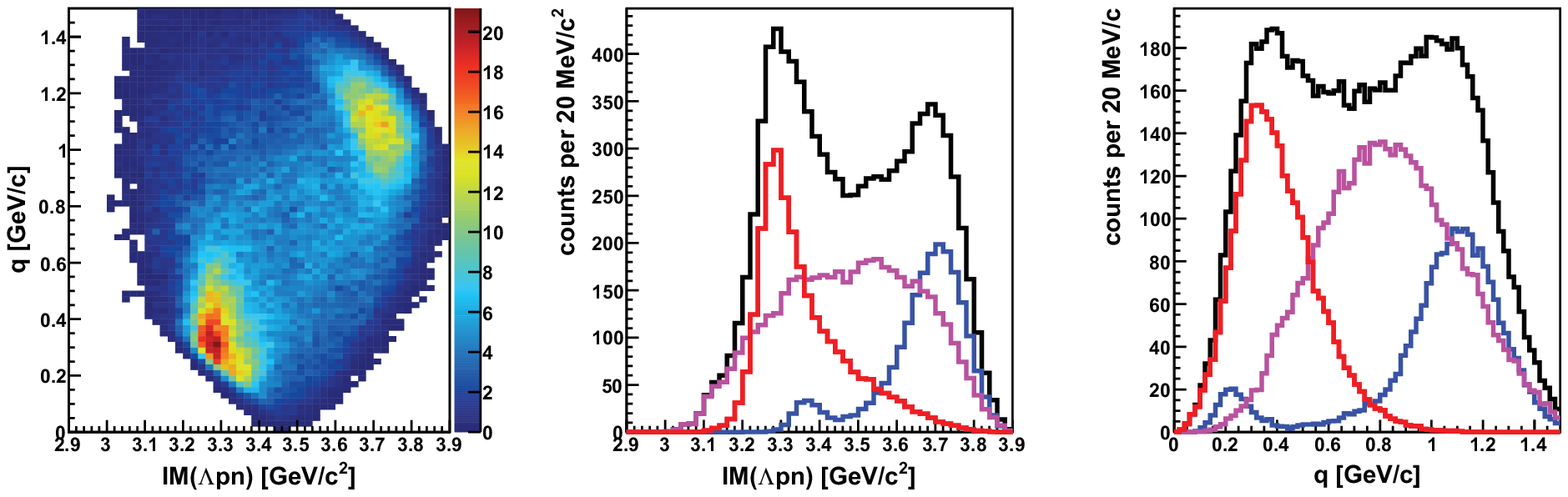}
  \caption{
  Expected spectra of (top) $\Lambda d$ and (bottom) 
  $\Lambda pn$ in the $K^- {^4}$He $\to \Lambda d+n$ and $\Lambda pn+n$ final
  state with 100-G kaons on target, respectively.
  We assume the cross section of the $K^-ppn$, quasi-free, and
  background to be 10 $\mu$b, 10 $\mu$b, and 20 $\mu$b, respectively, based
  on the E15 results.
  }
  \label{fig:spectra}
 \end{center}
\end{figure}

Figure~\ref{fig:spectra} shows the expected spectra of the $K^-$ $^4$He
$\to \Lambda d n$ and $\Lambda pnn$ final states with 100-G kaons on
target, corresponding to three weeks data taking under 90-kW beam power of the
J-PARC Main Ring accelerator~\cite{E80_proposal}.
For the $K^-ppn$ state in the figures, we assume a similar
distribution to that of $K^-pp$ observed in E15,
with a binding energy of $\sim$ 50 MeV, a width of $\sim$ 100 MeV, and
a Gaussian form factor of $\sim$ 400 MeV/$c$ with 10 $\mu$b ($\sigma
\cdot BR$).
For the quasi-free process and broad background, we also used 
similar parameters to those obtained for the $K^- {^3}$He $\to \Lambda pn$ final
state in E15.
The spectra for each process are generated using the Geant4 simulation,
in which we assume the same resolution of the existing CDS
to reconstruct the simulated tracks: $5.3\% \times p_t \oplus 0.5\%$
for charged particles~\cite{Sada:2016nkb} and $\sim 10\% \times p \oplus 7\%$ 
for a neutron~\footnote{based on the preliminary result of
$K^-$$^3{\rm He} \to \pi\Sigma pn$ analysis~\cite{Sakuma:2020dnc}}.

\begin{figure}[tbp]
 \begin{center}
  \includegraphics[width=0.4\textwidth,keepaspectratio]{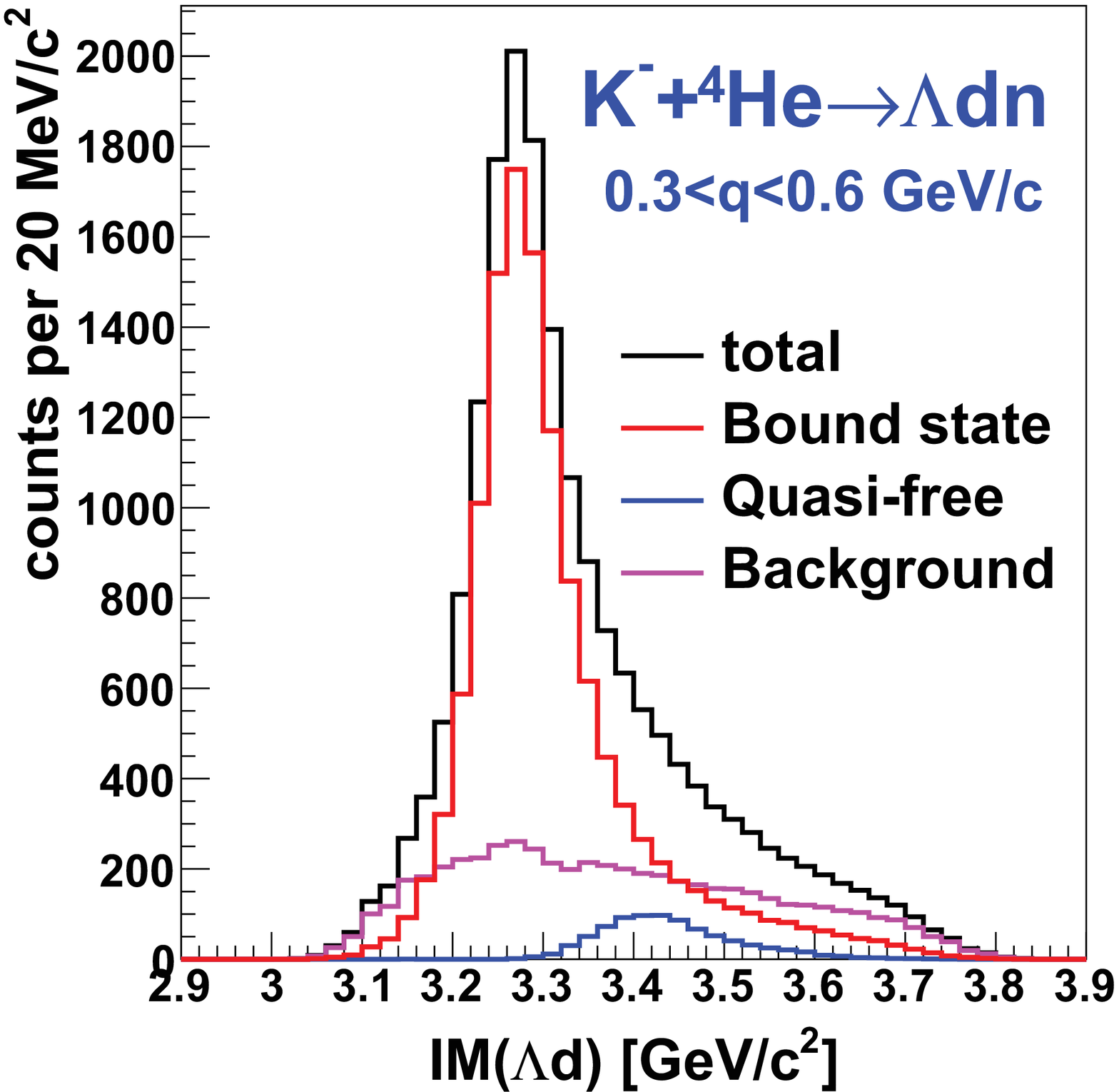}
  \includegraphics[width=0.4\textwidth,keepaspectratio]{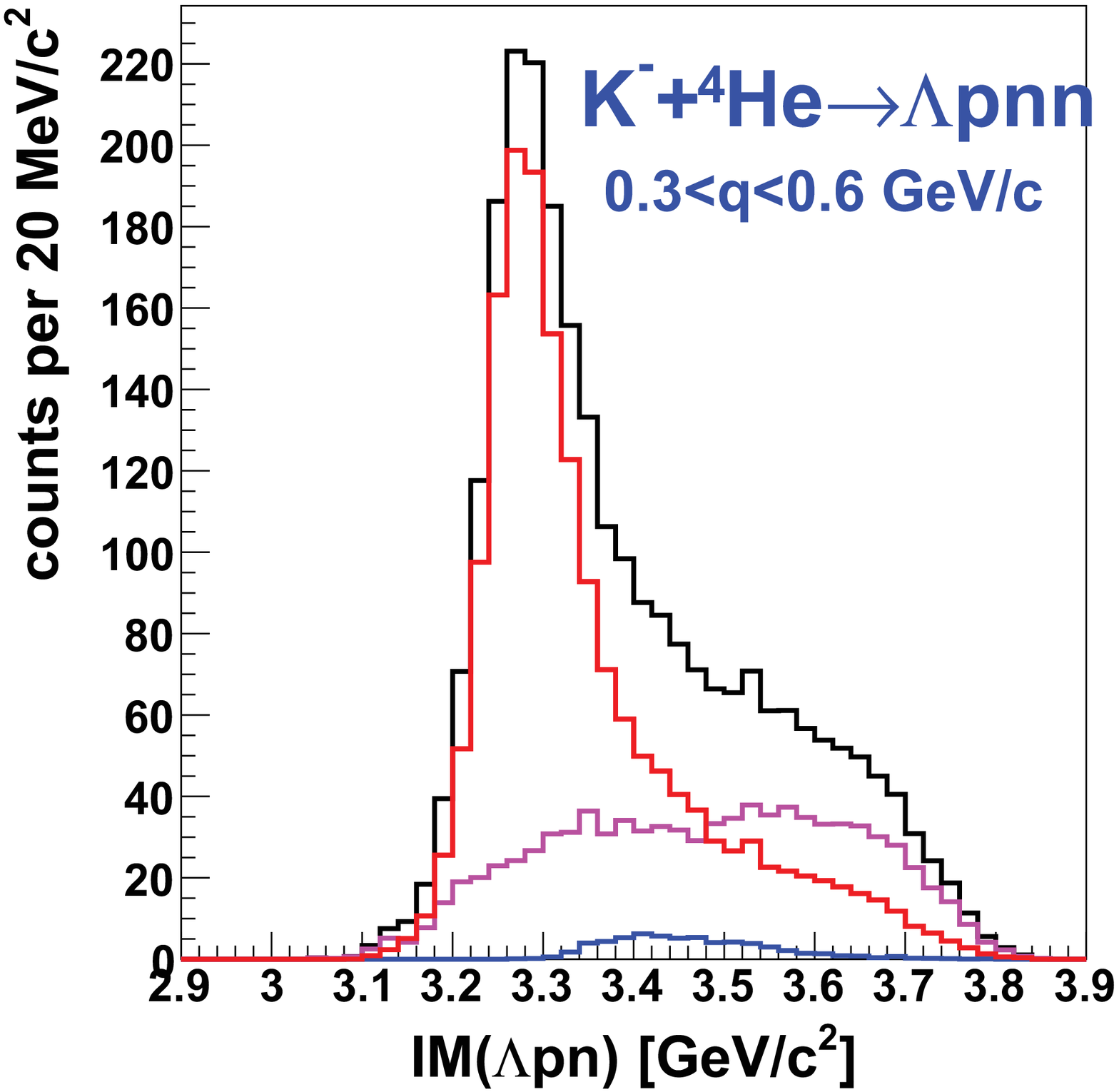}
  \caption{
  Expected invariant-mass spectra of (left) $\Lambda d$ and (light)
  $\Lambda pn$ in the region $0.3<q<0.6$ GeV/$c$, obtained
  from Fig.~\ref{fig:spectra}.
  }
  \label{fig:spectra2}
 \end{center}
\end{figure}

In the assumed conditions, the bound states are clearly identified not
only in the invariant mass spectra but also on the 2-dimensional plane of
the momentum transfer versus the invariant mass ($q$--$IM$) distribution,
which is the most important feature of the experiment.
By investigating the specific structures in the $q$--$IM$ distribution, we
can kinematically identify the signal of the bound state.
Figure~\ref{fig:spectra2} shows a demonstration of the signal
enhancement by selecting a momentum transfer of $0.3<q<0.6$ GeV/$c$.
In this region, we expect the bound state to be clearly separated from
the quasi-free process, as in the case of the E15 experiment shown in Fig.~\ref{fig:PLB}.

\section{Summary}
\label{summary}
We demonstrated that kaonic nuclei can be produced via in-flight
$(K^-,N)$ reactions using the low-momentum DC kaon beam at the J-PARC
E15 experiment.
We observed that the simplest kaonic nuclei, $K^-pp$, has a much deeper
binding energy than normal nuclei.
We also found that the large form factor obtained in a PWIA analysis
implies the possible formation of a compact and dense system.
For the next stage, we have proposed a series of experimental programs
for the systematic investigation of light kaonic nuclei, from 
$\bar K N$ ($\Lambda(1405)$) to $\bar K NNNN$.
Through the experiments, we will determine the features of kaonic
nuclei depending on the mass number $A$, {\it i.e.},
nuclear density, which is related to spontaneous and explicit chiral
symmetry breaking in QCD.
In the new experiment approved as J-PARC E80 we will measure
the $\bar K NNN$ ($A=3$) system as a first step toward a comprehensive
study. 

\begin{acknowledgements}
We are grateful to all the staff members of J-PARC/KEK/JAEA for
their extensive efforts in the successful operation of the facility. 
This work is supported in part by MEXT Grants-in-Aid 26800158, 17K05481,
26287057, 24105003, 14102005, 17070007, 18H05402, 18H01237, and 20K04006. 
Part of this work is also supported by the Ministero degli Affari Esteri e
della Cooperazione Internazionale, Direzione Generale per la Promozione
del Sistema Paese (MAECI), StrangeMatter project.
\end{acknowledgements}

\bibliographystyle{spphys.bst}       % APS-like style for physics
\bibliography{APFB2020}   % name your BibTeX data base

\providecommand{\noopsort}[1]{}\providecommand{\singleletter}[1]{#1}%
\begin{thebibliography}{10}
\providecommand{\url}[1]{{#1}}
\providecommand{\urlprefix}{URL }
\expandafter\ifx\csname urlstyle\endcsname\relax
  \providecommand{\doi}[1]{DOI \discretionary{}{}{}#1}\else
  \providecommand{\doi}{DOI \discretionary{}{}{}\begingroup
  \urlstyle{rm}\Url}\fi

\bibitem{Iwasaki:1997wf}
M.~Iwasaki, et~al., Phys. Rev. Lett. \textbf{78}, 3067 (1997)

\bibitem{Beer:2005qi}
G.~Beer, et~al., Phys. Rev. Lett. \textbf{94}, 212302 (2005)

\bibitem{Bazzi:2011zj}
M.~Bazzi, et~al., Phys. Lett. \textbf{B704}, 113 (2011)

\bibitem{Martin:1980qe}
A.D. Martin, Nucl. Phys. \textbf{B179}, 33 (1981)

\bibitem{Nogami:1963xqa}
Y.~Nogami, Phys. Lett. \textbf{7}, 288 (1963)

\bibitem{Akaishi:2002bg}
Y.~Akaishi, T.~Yamazaki, Phys. Rev. \textbf{C65}, 044005 (2002)

\bibitem{Yamazaki:2002uh}
T.~Yamazaki, Y.~Akaishi, Phys. Lett. \textbf{B535}, 70 (2002)

\bibitem{Shevchenko:2006xy}
N.V. Shevchenko, A.~Gal, J.~Mares, Phys. Rev. Lett. \textbf{98}, 082301 (2007)

\bibitem{Shevchenko:2007ke}
N.V. Shevchenko, A.~Gal, J.~Mares, J.~Revai, Phys. Rev. \textbf{C76}, 044004
  (2007)

\bibitem{Ikeda:2007nz}
Y.~Ikeda, T.~Sato, Phys. Rev. \textbf{C76}, 035203 (2007)

\bibitem{Dote:2008in}
A.~Dot{\'{e}}, T.~Hyodo, W.~Weise, Nucl. Phys. \textbf{A804}, 197 (2008)

\bibitem{Ikeda:2008ub}
Y.~Ikeda, T.~Sato, Phys. Rev. \textbf{C79}, 035201 (2009)

\bibitem{Wycech:2008wf}
S.~Wycech, A.M. Green, Phys. Rev. \textbf{C79}, 014001 (2009)

\bibitem{Dote:2008hw}
A.~Dot{\'{e}}, T.~Hyodo, W.~Weise, Phys. Rev. \textbf{C79}, 014003 (2009)

\bibitem{Ikeda:2010tk}
Y.~Ikeda, H.~Kamano, T.~Sato, Prog. Theor. Phys. \textbf{124}, 533 (2010)

\bibitem{Barnea:2012qa}
N.~Barnea, A.~Gal, E.Z. Liverts, Phys. Lett. \textbf{B712}, 132 (2012)

\bibitem{Bayar:2012hn}
M.~Bayar, E.~Oset, Phys. Rev. \textbf{C88}, 044003 (2013)

\bibitem{Maeda:2013}
S.~Maeda, Y.~Akaishi, T.~Yamazaki, Proc. Jpn. Acad. \textbf{B89}, 418 (2013)

\bibitem{Revai:2014twa}
J.~R\'evai, N.V. Shevchenko, Phys. Rev. \textbf{C90}, 034004 (2014)

\bibitem{Dote:2014via}
A.~Dot{\'{e}}, T.~Inoue, T.~Myo, Prog. Theor. Exp. Phys. \textbf{2015}, 043D02
  (2015)

\bibitem{Sekihara:2016vyd}
T.~Sekihara, E.~Oset, A.~Ramos, Prog. Theor. Exp. Phys. \textbf{2016}, 123D03
  (2016)

\bibitem{Ohnishi:2017uni}
S.~Ohnishi, et~al., Phys. Rev. \textbf{C95}, 065202 (2017)

\bibitem{Dote:2017veg}
A.~Dot{\'{e}}, T.~Inoue, T.~Myo, Phys. Rev. \textbf{C95}, 062201 (2017)

\bibitem{Dote:2017wkk}
A.~Dot{\'{e}}, T.~Inoue, T.~Myo, Phys. Lett. \textbf{B784}, 405 (2018)

\bibitem{Agnello:2005qj}
M.~Agnello, et~al., Phys. Rev. Lett. \textbf{94}, 212303 (2005)

\bibitem{Yamazaki:2010mu}
T.~Yamazaki, et~al., Phys. Rev. Lett. \textbf{104}, 132502 (2010)

\bibitem{Ichikawa:2014ydh}
Y.~Ichikawa, et~al., Prog. Theor. Exp. Phys. \textbf{2015}, 021D01 (2015)

\bibitem{Doce:2015ust}
O.~Vzquez~Doce, et~al., Phys. Lett. \textbf{B758}, 134 (2016)

\bibitem{DelGrande:2018sbv}
R.~Del~Grande, et~al., Eur. Phys. J. \textbf{C79}, 190 (2019)

\bibitem{Agakishiev:2014dha}
G.~Agakishiev, et~al., Phys. Lett. \textbf{B742}, 242 (2015)

\bibitem{Tokiyasu:2013mwa}
A.O. Tokiyasu, et~al., Phys. Lett. \textbf{B728}, 616 (2014)

\bibitem{Hashimoto:2014cri}
T.~Hashimoto, et~al., Prog. Theor. Exp. Phys. \textbf{2015}, 061D01 (2015)

\bibitem{Sada:2016nkb}
Y.~Sada, et~al., Prog. Theor. Exp. Phys. \textbf{2016}, 051D01 (2016)

\bibitem{Ajimura:2018iyx}
S.~Ajimura, et~al., Phys. Lett. \textbf{B789}, 620 (2019)

\bibitem{J-PARCE15:2020gbh}
T.~Yamaga, et~al., Phys. Rev. C \textbf{102}, 044002 (2020)

\bibitem{KKpp_LoI}
{\rm Letter of Intent for J-PARC, 2009, "Double Anti-kaon Production in Nuclei
  by Stopped Anti-proton Annihilation"},
  http://j-parc.jp/researcher/Hadron/en/Proposal\_e.html

\bibitem{Sekihara:QNP2018}
T.~Sekihara, E.~Oset, A.~Ramos, JPS Conf. Proc. \textbf{26}, 023009 (2019)

\bibitem{Sato:2007sb}
M.~Sato, et~al., Phys. Lett. B \textbf{659}, 107 (2008)

\bibitem{Yim:2010zza}
H.~Yim, et~al., Phys. Lett. B \textbf{688}, 43 (2010)

\bibitem{Agnello:2007aa}
M.~Agnello, et~al., Phys. Lett. B \textbf{654}, 80 (2007)

\bibitem{Herrmann:2005}
N.~Herrmann, et~al., Proc. EXA05 Conference \textbf{ISBN 3-7001-3616-1}, 73
  (2005)

\bibitem{E80_proposal}
{\rm Proposal for J-PARC, 2020, "Systematic investigation of the light kaonic
  nuclei"}, http://j-parc.jp/researcher/Hadron/en/Proposal\_e.html

\bibitem{Sakuma:2020dnc}
F.~Sakuma, et~al., AIP Conf. Proc. \textbf{2249}, 020005 (2020)

\end{thebibliography}

%\begin{thebibliography}{}
% etc
%\end{thebibliography}

\end{document}